\documentclass{article}
\usepackage{fortschritte}
\usepackage{amsmath, amssymb, amstext}

\def\({\left(}
\def\){\right)}

\newcommand{\Exc}[1]{{${\rm E}_{{#1}({#1})}$}}

\def\mxth{\mathsurround=0pt }
\def\xversim#1#2{\lower2.pt\vbox{\baselineskip0pt \lineskip-.5pt
x  \ialign{$\mxth#1\hfil##\hfil$\crcr#2\crcr\sim\crcr}}}

\renewcommand{\a}{\alpha}
\renewcommand{\b}{\beta}

\renewcommand{\d}{\delta}

\newcommand{\g}{\gamma}

\newcommand{\m}{\mu}
\newcommand{\n}{\nu}

\def\be{\begin{equation}}
\def\ee{\end{equation}}
\def\bea{\begin{eqnarray}}
\def\eea{\end{eqnarray}}

\newcommand{\ft}[2]{{\textstyle\frac{#1}{#2}}}

\newcommand{\eqn}[1]{(\ref{#1})}

\begin{document}
\def\titleline{Gauging Maximal Supergravities}

\def\email_speaker{
{\tt 
B.deWit@phys.uu.nl}}

\def\authors{Bernard de Wit\1ad\sp, Henning Samtleben\1ad and Mario
Trigiante\2ad} 

\def\addresses{
\1ad{Institute for Theoretical Physics and Spinoza Institute,\\
  Utrecht University, The Netherlands}\\
\2ad{Dipartimento di Fisica, Politecnico di Torino, Torino, Italy}\\
}

\def\abstracttext{
We review recent progress in the gauging of maximal supergravity
theories.} 

\large
\makefront
\section{Introduction}
Gaugings are the only known supersymmetric deformations of maximal
supergravity. They may originate in various ways from fluxes and
branes in higher dimensions. A gauging is obtained by coupling the
abelian vector fields, which arise in toroidally compactified
eleven-dimensional or IIB supergravity, to charges assigned to the
elementary fields. The resulting gauge group is encoded in these
charges, but supersymmetry severely restricts the possible gauge groups. 

Originally, gaugings of maximal supergravity theories were constructed
for gauge groups whose existence could be inferred from a
Kaluza-Klein interpretation. The first examples were ${\rm SO}(8)$ in
four dimensions \cite{dWN82} and ${\rm SO}(5)$ in seven dimensions
\cite{PPvN84}, related to $S^7$ and $S^4$ compactifications of
eleven-dimensional supergravity, respectively,  and later ${\rm SO}(6)$ in 
five dimensions \cite{GRW86}, related to the $S^5$ compactification of
IIB supergravity. At the same time it was also demonstrated how
noncompact versions of these orthogonal gauge groups and contractions
thereof could lead to viable gaugings \cite{Hull84}. After these
initial developments the subject lay dormant for quite some time,
until the importance of anti-de Sitter spaces, which are the natural
ground states for gauged supergravities, became apparent. In the last
few years new gaugings were discovered and explored (see, e.g. 
\cite{collect}), motivated by the adS/CFT correspondence and by the
study of flux compactifications. Therefore it is a timely question to
reinvestigate these gaugings from a more general viewpoint. 

In this paper we review recent progress in this direction
\cite{dWST1,dWST2}. 
A central feature of the maximal supergravities is that the scalars
parametrize a ${\rm G}/{\rm H}$ symmetric space. Here, ${\rm G}$ is
usually referred to as the `duality group' and ${\rm H}$ coincides
with the R-symmetry group. 
The scalar fields are then described in terms of a spacetime dependent
element of ${\rm G}$, denoted by ${\cal V}(x)$, which transforms
under rigid ${\rm G}$ transformations from the left and under local
${\rm H}$ transformations from the right. Upon choosing a gauge, the
group element ${\cal V}$ becomes the coset representative of ${\rm
G}/{\rm H}$. Because ${\rm H}$ is already realized as a local
invariance (with composite gauge fields), a gauging of the
supergravity theory must be effected by embedding the new gauge group
associated with the elementary gauge fields into the duality group 
${\rm G}$. The embedding of this gauge group into ${\rm G}$ is
described by an embedding tensor $\Theta_M{}^\alpha$, which we will
introduce in the next section. What we want to stress here is the
following. In the gauging the kinetic terms are modified by 
covariantizing spacetime derivatives and field strengths with respect
to the gauge group, without altering their appearance in the
Lagrangian. In the same way the masslike terms and  
the potential associated with the gauging are described by the
so-called $T$-tensor \cite{dWN82}, which appears in the Lagrangian and
the supersymmetry transformations in a way that is independent of the
specific gauge group. Of course, just as the covariant derivatives
and the field strengths, the $T$-tensor depends on the embedding
tensor, but this dependence is implicit.

\section{The embedding tensor}
As explained above, the central question is which gauge groups can be
embedded into the duality group ${\rm G}$ such that the
supersymmetry of the Lagrangian can be preserved. The latter requires
to introduce additional terms to the action and the supersymmetry
transformation rules, which take a uniform form in terms of the
$T$-tensor. In this section we discuss the embedding of the gauge group.

The (abelian) vector fields $A_\m^M$ transform in a representation of
the duality group ${\rm G}$ with generators denoted by
$(t_\a)_M{}^N$, so that $\d A_\m{}^M = -\Lambda^\a (t_\a)_N{}^M\,A_\m{}^N$.  
The gauging is effected by introducing 
gauge group generators $X_M$, which couple to the gauge fields and
depend linearly on the ${\rm G}$-generators $t_\a$, {\it i.e.},
\be
\label{X-theta-t}
X_M = \Theta_M{}^\a\,t_\a\;,
\ee
so that the gauging is characterized by a real {\it embedding tensor}
$\Theta_{M}{}^{\alpha}$. The fact that the $X_M$ generate a group, 
implies that the embedding tensor satisfies the closure condition,
\be
\label{gauge-gen}
\Theta_M{}^\a\,\Theta_N{}^\b \,f_{\a\b}{}^{\g}= f_{MN}{}^P\,
\Theta_P{}^\g\,,
\ee
where the $f_{\a\b}{}^\g$ and $f_{MN}{}^P$ are the structure constants
of ${\rm G}$ and of the gauge group, respectively. This condition
implies that $\Theta$ is invariant under the gauge group. 
It is possible to rewrite the right-hand side of \eqn{gauge-gen} in
terms of the generators $t_\a$, so that 
\be
\label{closure-constraint}
f_{\b\g}{}^\a \, \Theta_M{}^\b\,\Theta_N{}^\g - (t_\b)_N{}^P\,
\Theta_M{}^\b \, \Theta_P{}^\a =0 \;.
\ee
In four spacetime dimensions, the situation is somewhat different as
one is dealing with both electric and 
magnetic charges, which together constitute the fundamental
representation of ${\rm G}= {\rm E}_{7(7)}$. Here the
gauge fields do not transform under the full duality group, as the
dual magnetic potentials are lacking in the Lagrangian. The embedding 
tensor must therefore vanish for magnetic charges. We return to this
issue in sect.~4. 

The $T$-tensor is the ${\rm H}$-covariant, field-dependent, tensor
defined  by 
\be
\label{T-theta}
T_M{}^\a[\Theta,\phi]\,t_\a  = {\cal V}^{-1}{}_{\!\!\!\!M}{}^N\,
\Theta_N{}^\a\,  ({\cal
V}^{-1}t_\a  {\cal V} )\;,
\ee
where ${\cal V}$ denotes the coset representative of ${\rm G}/{\rm
H}$. When treating the embedding tensor as a spurionic object that 
transforms under the duality group, the Lagrangian and 
transformation rules remain formally invariant under ${\rm G}$. Under
such a transformation $\Theta$ would transform as $\Theta_M{}^\a\, t_\a \to
g_M{}^N\,\Theta_N{}^\a\,(g\,t_\a g^{-1})$, with $g\in {\rm G}$. Of
course, when freezing $\Theta_M{}^\a$ to a constant, the ${\rm
G}$-invariance is broken. 

Because the closure identity \eqn{closure-constraint} is covariant
with respect to ${\rm G}$, it leads to a corresponding ${\rm
H}$-covariant constraint quadratic in the $T$-tensor. Subsequent
considerations will show that there is also a constraint linear in the
$T$-tensor. This tensor contains a number of ${\rm H}$-covariant
tensors conventionally denoted  
by $A_1$, $A_2$ and $A_3$, which appear in the fermionic masslike
terms proportional to $\psi_\m\psi_\n$, $\psi_\m\chi$ and
$\chi\chi$, respectively. Here $\psi_\m$ denotes the gravitini
fields and $\chi$ the matter spinor fields, whose supersymmetry
variations $\d\psi_\m$ and $\d\chi$ acquire terms linear in $A_1$ and
$A_2$, respectively. In addition to the masslike terms the Lagrangian
contains a scalar potential quadratic in $A_1$ and in $A_2$. 
Given the fact that gaugings are the only possible supersymmetric
deformations of maximal supergravity, the $T$-tensor cannot contain
any other tensors beyond $A_{1,2,3}$. Because the fermion fields 
transform according to known representations of ${\rm H}$, the
representation content of $A_{1,2,3}$ is in fact determined uniquely.

\begin{table}
\begin{center}
\begin{tabular}{l l  l l  }\hline
~&~&~&~\\[-4mm]
$d$ &${\rm G}$& ${\rm H}$ & $T$  \\   \hline
~&~&~&~\\[-4mm]
7   & ${\rm SL}(5)$ & ${\rm USp}(4)$  & ${\bf 10}\times {\bf 24}= {\bf
  10}+\underline{\bf 15}+  {\bf 40}+ {\bf 175}$  \\[1mm]
6  & ${\rm SO}(5,5)$ & ${\rm USp}(4) \times {\rm USp}(4)$ & 
  ${\bf 16}\times{\bf 45} =
  {\bf 16}+ \underline{\bf 144} + {\bf 560}$ \\[.8mm]
5   & ${\rm E}_{6(6)}$ & ${\rm USp}(8)$ & ${\bf 27}\times{\bf 78} = 
  {\bf 27} + \underline{\bf 351} + {\bf 1728}$  \\[.5mm]
4   & ${\rm E}_{7(7)}$ & ${\rm SU}(8)$  & ${\bf 56}\times{\bf 133} = 
  {\bf 56} + \underline{\bf 912} + {\bf 6480}$   \\[.5mm]
3   & ${\rm E}_{8(8)}$ & ${\rm SO}(16)$ & ${\bf 248}\times{\bf 248} = 
  \underline{\bf 1} + {\bf 248} + \underline{\bf 3875} +{\bf 27000} 
  +  {\bf 30380}$ 
\\ \hline
\end{tabular}
\end{center}
\caption{\small
Decomposition of the $T$-tensor in various
dimensions for maximal supergravities in terms of irreducible
representations of G. According to the representation constraint, only
the underlined representations are allowed.
}\label{T-tensor-repr}
\end{table}

Now we note that {\it every} variation of the coset representative can
be expressed as a (possibly field-dependent) ${\rm G}$-transformation
acting on ${\cal V}$ from the right. For example, a rigid duality
transformation acting from the left, 
can be rewritten as a field-dependent transformation from the right, 
\be
{\cal V}\to {\cal V}^\prime = g\,{\cal V}= {\cal V} \, \sigma^{-1}\,. 
\ee
with $\sigma^{-1}= {\cal V}^{-1}\,g\,{\cal V}\in {\rm G}$, but also a
supersymmetry transformation can be written in this
form. Consequently, these 
variations of ${\cal V}$ induce the following transformation of the
$T$-tensor, 
\be
T_M{}^\a\, t_\a \to T^{\prime}_M{}^\a\,t_\a = 
\sigma_M{}^N\,T_N{}^\a\,(\sigma\, t_\a \sigma^{-1})\;.
\ee
This implies that the $T$-tensor must constitute a
representation of ${\rm G}$. Observe that this is {\it not} an invariance
statement; rather it means that the
$T$-tensor (irrespective of the choice for the corresponding embedding
tensor) varies under supersymmetry or any other transformation in a
way that can be written as a (possibly field-dependent) ${\rm
G}$-transformation. Note also that the transformation assignment of
the embedding tensor and the $T$-tensor are opposite in view of the
relationship between $g$ and $\sigma$.  Since we know that the
$T$-tensor representation should be decomposable into the
${\rm H}$-representations associated with $A_{1,2,3}$,we can almost
uniquely identify the representation to which the $T$-tensor 
belongs. As it turns out this leads to a restricted ${\rm
G}$-representation for the $T$-tensor, and therefore for the embedding
tensor. The result of this analysis is displayed in table~1 for
spacetime dimensions $d=3,\ldots,7$. 

Hence, consistent gaugings are characterized by embedding tensors that
satisfy two constraints, one quadratic and one linear in the embedding
tensor. The quadratic constraint
ensures that the embedding tensor defines a proper subgroup of the
duality group. The linear constraint implies that the embedding
tensor belongs to a specific representation of the duality group,
so that the corresponding $T$-tensor matches precisely with the
tensors $A_1$, $A_2$ and $A_3$ that appear in the fermionic masslike
terms in the Lagrangian.

\section{Example: some gaugings in $d=5$ dimensions}
As an application, let us consider some of the gaugings in $d=5$
maximal supergravity and search for viable gauge groups embedded in
the \Exc6 duality group. We first assume that the gauge group is a subgroup
of the  ${\rm SL}(2,\mathbb{R})\times {\rm SL}(6,\mathbb{R})$ maximal
subgroup of \Exc6. According to table~1, the embedding tensor must
belong to the $({\bf 27}\times{\bf 78})\cap{\bf 351}$
representation. With respect to 
${\rm SL}(2,\mathbb{R})\times {\rm SL}(6,\mathbb{R})$, the vector
gauge fields, the \Exc6 generators and the embedding tensor decompose
according to,  
\begin{eqnarray}
\overline{\bf 27} &\rightarrow& ({\bf 1},\overline{{\bf 15}}) +
({\bf 2},{\bf 6}) \,,
\nonumber\\
{\bf 78} &\rightarrow&
({\bf 1},{\bf 35}) + ({\bf 3},{\bf 1})
+ ({\bf 2},{\bf 20})\,,
\nonumber\\
{\bf 351} &\rightarrow&
({\bf 1},{\bf 21}) + ({\bf 3},{\bf 15}) + ({\bf 2},\overline{\bf 84})
+ ({\bf 2},\overline{\bf 6}) + ({\bf 1},{\bf 105})\,,
\label{decblock-sl6}
\end{eqnarray}
respectively.
The table below summarizes how the embedding tensor couples the vector
fields to the generators, 
\begin{equation}
\begin{tabular}{|c| c c| } \hline
~&~&~\\[-4.5mm]
&$({\bf 1},{\bf 15})$
&$({\bf 2},\overline{\bf 6})$
\\ \hline
~&~&~\\[-4.5mm]
$({\bf 1},{\bf 35})$
&$({\bf 1},{\bf 21}) + ({\bf 1},{\bf 105}) $
&$({\bf 2},\overline{\bf 6})+({\bf 2},\overline{\bf 84}) $
\\ 
$({\bf 3},{\bf 1})$
&$({\bf 3},{\bf 15})$
&$({\bf 2},\overline{\bf 6})$
\\ 
$({\bf 2},{\bf 20})$
&$({\bf 2},\overline{\bf 6})+({\bf 2},\overline{\bf 84}) $
&$({\bf 3},{\bf 15}) + ({\bf 1},{\bf 105}) $
\\ \hline
\end{tabular}
\end{equation}
Observe that the left column refers to the adjoint representation of
\Exc6 written with upper index $\a$ as it appears in $\Theta_M{}^\a$,
whereas the decomposition of the gauge group generators 
\eqn{X-theta-t} refers to the \Exc6 generators $t_\a$ written with lower
index $\a$. The top line specifies the possible charges to which the 
gauge fields can couple, which transform in the conjugate
representations as compared to the gauge fields. In view of the fact
that all representations in the 
${\bf 351}$ appear with multiplicity 1, equivalent representations in
the table must be identified. Because we assume that the gauge group
is contained in ${\rm SL}(2,\mathbb{R})\times {\rm SL}(6,\mathbb{R})$,
there is only one possible representation assignment
for the embedding tensor,  namely it should belong to the $({\bf
1},{\bf 21})$ representation, which does not appear in the
bottom row. Hence, only the vector fields transforming
in the $({\bf 1},\overline{\bf 15})$ representation are involved in
the gauging and couple to the generators in the adjoint representation
of ${\rm SL}(6,\mathbb{R})$. The vector fields in the $({\bf 2},{\bf
6})$ cannot participate in the gauging and must be dualized into 
charged massive antisymmetric tensor fields. 

Because the embedding tensor belongs to the $({\bf 1},{\bf 21})$
representation, its nonzero components are parametrized in terms 
of a six-by-six symmetric tensor $\theta_{AB}$ according to
$\Theta_{[AB]}{}^{\!C}{}_{\!D} = \d^C_{[A}\,\theta^{~}_{B]D}$, where
$\theta_{AB}$ is characterized by $p$ eigenvalues $+1$, $q$
eigenvalues $-1$ and $r=6-p-q$ eigenvalues 0. The gauge group
generators can be decomposed in terms of the ${\rm SL}(6,\mathbb{R})$
generators  $t_A{}^{\!B}$ as follows, 
\be 
X_{AB}= \Theta_{[AB]}{}^{\!C}{}_{\!D}  \; t_C{}^{\!D} 
= \theta^{~}_{D[A} \,t_{B]}{}^{\!D} \,.
\ee
These generators define the ${\rm CSO}(p,q,r)$ subgroup of ${\rm
SL}(6,\mathbb{R})$, which has dimension 
$15-\ft12r(r-1)$ and leaves the embedding tensor 
invariant. The gaugings are thus completely
determined and encoded in the 27 conjugacy classes of $\theta_{AB}$
leading to 15 inequivalent 
gaugings \cite{GRW86,AndCordFreGual}. 
Because we already established the closure of the gauge group,
the quadratic constraint \eqn{closure-constraint} on the embedding
tensor does not give rise to additional restrictions.

A second application is based on the subgroup ${\rm SO}(5,5)\times {\rm
SO}(1,1)$. This semisimple group is not a maximal subgroup of \Exc6,
but it becomes maximal upon including 16 additional nilpotent
generators transforming in the  ${\bf 16}_{-3}$
representation. We consider gauge groups that are a subgroup of
this non-semisimple maximal subgroup. 
The decomposition of the
relevant ${\rm E}_{6(6)}$ representations with respect to 
${\rm SO}(5,5)\times {\rm SO}(1,1)$ is given by,
\begin{eqnarray}
\overline{\bf 27} &=& \overline{\bf 16}_{-1}
+{\bf 10}_{+2}+{\bf 1}_{-4}\,,
\nonumber\\
{\bf 78} &=& {\bf 45}_{0}+{\bf 1}_{0}+ {\bf 16}_{-3}+\overline{{\bf
16}}_{+3}\,,\nonumber\\
{\bf 351} &=& {\bf 144}_{+1}+
{\bf 16}_{+1}+{\bf 45}_{+4}+{\bf 120}_{-2}+{\bf 10}_{-2}+{\bf
\overline{16}}_{-5}\,.
\end{eqnarray}
The couplings induced by the embedding tensor are shown in the table below, 
\begin{equation}
\label{5,5}
\begin{tabular}{|c|ccc|} \hline
~&~&~&~\\[-4.5mm]
&${\bf 16}_{+1}$
&${\bf 10}_{-2}$
&${\bf 1}_{+4}$
\\
\hline
~&~&~&~\\[-4.5mm]
${\bf 45}_{0}$
& ${\bf 144}_{+1}+ {\bf 16}_{+1}$
&${\bf 10}_{-2}+{\bf 120}_{-2}$
& ${\bf 45}_{+4}$
\\
${\bf 1}_{0}$
&${\bf 16}_{+1}$
&${\bf 10}_{-2}$
&
\\
${\bf 16}_{-3}$
&${\bf 120}_{-2}+ {\bf 10}_{-2}$
&${\overline{\bf 16}}_{-5} $
&${\bf 16}_{+1}$
\\
$\overline{{\bf16}}_{+3}$
&${\bf 45}_{+4}$
&${\bf 144}_{+1}+ {\bf 16}_{+1}$
&\\
\hline
\end{tabular}
\end{equation}
Again equivalent representations in the embedding matrix should
be identified as they appear with multiplicity one in the ${\bf 351}$
representation. The generators transforming as the ${\bf
16}_{-3}$ representation (denoted in the table above by 
the conjugate $\overline{\bf 16}_{+3}$) cannot be involved in the
gauging, as they 
do not belong  to the maximal subgroup that we have selected. Therefore
only two representations are allowed for the embedding tensor, namely
${\bf 144}_{+1}$ and ${\bf 45}_{+4}$. 
No gaugings have been worked out so far with embedding
tensors that transform reducibly as ${\bf 144}+{\bf 45}$. The two
irreducible cases can readily be identified and may originate from $d=6$
dimensions.  An embedding tensor belonging to the ${\bf 144}_{+1}$ 
representation is induced by $d=6$ gauged supergravity, as its 
embedding matrix must belong to the ${\bf 144}$ 
representation of the ${\rm SO}(5,5)$ duality group ({\it c.f.}
table~1), upon dimensional reduction on $S^1$. 

An embedding tensor belonging to the ${\bf 45}_{+4}$
representation is obtained by a Scherk-Schwarz reduction from
$d=6$ dimensions, where ungauged maximal supergravity is invariant
under ${\rm SO}(5,5)$ duality. Indeed, the representations of 
the vector fields are in accord with this interpretation. 
The embedding tensor is parametrized in terms of a matrix
$\theta_p{}^q$, with $p,q=1,\ldots ,16$, belonging to the spinor 
representation of ${\rm SO}(5,5)$, such that
\begin{eqnarray}
X_0&=&\theta_p{}^q\, t^p{}_q \,, \qquad  X_p = {\theta}_p{}^q \, t_q\,,
\label{embefla5}
\end{eqnarray}
where the $t^p{}_q$ are the generators of ${\rm SO}(5,5)$, while $t_p$
are the generators belonging to the ${\bf 16}_{-3}$
representation. This differs from the assignment in the
table~(\ref{5,5}), because the embedding tensor has an upper index
$\a$ associated with the Lie algebra of \Exc6, whereas the generators
carry lower indices. The gauge algebra obviously closes,
\begin{eqnarray}
\label{55-algebra}
{[X_0,\,X_p]}= {\theta}_p{}^q X_q\,, \qquad [X_p,\,X_q]=0\,.
\end{eqnarray}
{From} (\ref{embefla5}) it follows that the null vectors of
$\theta_p{}^q$ correspond to gauge fields that do not participate in
the gauging and 
remain abelian. Obviously the maximal dimension of the gauge group is
equal to $17$. The vector fields in the ${\bf 10}_{+2}$ are
generically charged under $X_0$ and must be converted to
charged tensor fields in order to be described in terms of a
Lagrangian. Because the gauge group closes in view of  
\eqn{55-algebra}, the quadratic constraint \eqn{closure-constraint}
must be satisfied, so that there are no further restrictions. 

The above examples have consistent gauge groups, so there is no need
to verify the quadratic constraint. This is no longer obvious when
choosing the embedding tensor in a reducible representation, or when
assuming that the gauge group is embedded in the maximal
compact subgroup generated by the generators of 
${\rm SO}(5,5)\times {\rm SO}(1,1)$ combined with the generators
transforming in the $\overline{\bf 16}_{+3}$ representation. 
In this case there are many more options for the embedding tensor
consistent with the representation constraint. A more complete
analysis is far from trivial and should involve the quadratic
constraint \eqn{closure-constraint}. 

\section{Subtleties in $d=4$ dimensions}
In four spacetime dimensions the Lagrangian is not invariant under
${\rm G}={\rm E}_{7(7)}$, 
although the combined field equations and the Bianchi identities are. 
The Lagrangian is invariant under a subgroup ${\rm G}_{\rm
electric}\subset {\rm E}_{7(7)}$ and the gauge group has to be a
subgroup of ${\rm G}_{\rm electric}$. The Lagrangian is not unique as
there are many Lagrangians leading to equivalent field equations and
Bianchi identities, each one with a corresponding invariance
group ${\rm G}_{\rm electric}$. For any given Lagrangian gaugings
can be studied along the lines presented earlier. As an example we
consider the Lagrangian with ${\rm G}_{\rm electric}= {\rm
SL}(8,\mathbb{R})$. Hence we start with the branching rules under this
group of the relevant representations,
\begin{eqnarray}
{\bf 56}&\rightarrow & {\bf 28}+\overline{\bf 28}\,, \nonumber\\
{\bf 133}&\rightarrow & {\bf 63}+{\bf 70}\,, \nonumber\\
{\bf 912}&\rightarrow & {\bf 36}+{\bf 420}+ \overline{\bf 36}+
\overline{\bf 420} \,,
\label{sl8decs}
\end{eqnarray}
where the ${\bf 28}$ representation in the first branching corresponds
to the gauge potentials and the conjugate $\overline{\bf 28}$
corresponds to the dual magnetic potentials, which cannot be
incorporated in the gauging.  The branchings of products of the
relevant   
representations (\ref{sl8decs}) that belong to the ${\bf 912}$, and
thus identify acceptable representions of a $T$-tensor, is
conveniently summarized by the table below, 
\begin{equation}
\begin{tabular}{|c|cc|} \hline
~&~&~\\[-4.5mm]
&${\bf 28}$ &$\overline{\bf 28}$
\\
\hline
~&~&~\\[-4.5mm]
${\bf 63}$
&${\bf 36}+{\bf 420}$
&$\overline{\bf 36}+\overline{\bf 420}$
\\
${\bf 70}$
&$\overline{\bf 420}$
&${\bf 420}$
\\
\hline
\end{tabular}
\label{decblock}
\end{equation} 
The ${\bf420}$ and the $\overline{\bf420}$ representations appear
twice in the above table but have multiplicity~1 according to
\eqn{sl8decs}. Therefore their presence implies a coupling to both 
electric and magnetic charges, which is not
permitted. That leaves an 
embedding tensor transforming in the $\overline{\bf 36}$ 
representation as the only possibility. According to (\ref{decblock}),
the gauge group generators are then decomposable in the generators of 
${\rm SL}(8,\mathbb{R})$. Thus we conclude that all possible gaugings
for the Lagrangian in the ${\rm SL}(8,\mathbb{R})$ basis are 
encoded by an embedding matrix in the $\overline{\bf 36}$
representation. The gauging is completely determined 
and encoded in the 44 nontrivial conjugacy classes of an eight-by-eight
symmetric tensor transforming in the  $\overline{\bf 36}$ representation,
characterized by its eigenvalues $\pm1$ or
$0$. These conjugacy classes correspond to 24 inequivalent gaugings
with gauge group ${\rm CSO}(p,q,r)$ (this time with $p+q+r=8$) and
dimension $28 -\ft12r(r-1)$. As we are dealing with a consistent gauge
group, the quadratic constraint \eqn{closure-constraint} becomes
superfluous, as in the previous cases. 

Because the \Exc7 charges combine electric and magnetic ones, there is
a new feature in this case. The requirement that the charges can all
be chosen as electric ones (upon a suitable electric-magnetic duality
transformation) so that the gauge group can be embedded into a
subgroup ${\rm G}_{\rm electric}$ of a certain Lagrangian, implies
that the embedding tensor should satisfy an \Exc7-invariant quadratic
constraint, 
\be
\label{quadratic-constraint} 
\Theta_M{}^\a \,\Theta_N{}^\b \; \Omega^{MN}=0\;,
\ee 
where $\Omega^{MN}$ is the \Exc7 invariant symplectic matrix. 
This constraint has been proven directly for the $d=4$ $T$-tensor
based on electric charge assignments \cite{dWST1}, and it may
be compared to the previous quadratic constraint
\eqn{closure-constraint}. In order 
to do this, consider the symmetric product of two ${\bf 912}$
representations, 
\be
({\bf 912}\times {\bf 912})_{\rm symmetric}=  {\bf 133}+{\bf 8645}+
{\bf 1463}+{\bf 152152}+{\bf 253935}\;, 
\ee
It turns out that the constraint \eqn{quadratic-constraint} constitutes
precisely the first two representations, ${\bf 133}+ {\bf
8645}$. Furthermore, the constraint \eqn{closure-constraint}
constitutes the same representations, provided the embedding tensor is
restricted to the ${\bf 912}$ representation. Hence, 
\eqn{quadratic-constraint} is therefore {\it not}
independent. However, this situation leads to the 
observation that, when 
imposing the representation constraint that requires the embedding
tensor to belong to the ${\bf 912}$ representation, one can follow two
different strategies. Either one keeps track of the assignments of 
electric and magnetic charges (as we did above), in which case the
quadratic constraint \eqn{closure-constraint} is automatically
satisfied, or, one directly imposes the constraint
\eqn{closure-constraint}, in which case it is guaranteed that there
will exist a suitable Lagrangian (via electric-magnetic  
duality) such that the corresponding gauging can be switched on. The
latter is the strategy that we follow in the application described
below.   

\section{Gaugings from IIB fluxes}
As a last example we consider gaugings of
maximal supergravity in four spacetime dimensions that can in
principle be generated by three- and five-form fluxes of the type-IIB
theory. The proper setting is based on the decomposition of the \Exc7 
group according to ${\rm SL}(2,\mathbb{R})\times {\rm
SL}(6,\mathbb{R})$. The relevant 
embedding proceeds as follows, 
\bea
{\rm E}_{7(7)} \longrightarrow  {\rm SL}(6,\mathbb{R})\times {\rm
SL}(3,\mathbb{R})
\longrightarrow {\rm SL}(6,\mathbb{R})\times{\rm SL}(2,\mathbb{R})
\times {\rm SO}(1,1) \;.
\eea
There is another, inequivalent, embedding, but the one above is
relevant for the IIB theory, with ${\rm SL}(2,\mathbb{R})$ the
S-duality group. Under this embedding, the ${\bf 56}$ representation
of electric and magnetic charges, and the adjoint representation of
\Exc7 decompose as follows,  
\bea
\label{sl(6/2)-decomposition}
{\bf 56} &\to & 
(\overline{\bf 6}, {\bf 1})_{-2} + ({\bf 6}, {\bf 2})_{-1} 
 + ({\bf 20}, {\bf 1})_0 +
(\overline{\bf 6}, {\bf 2})_{+1} +({\bf 6},{\bf 1})_{+2}\;,  
\nonumber \\ 
{\bf 133} &\to & 
({\bf 1},{\bf2})_{-3}+ ({\bf 15},{\bf 1})_{-2}+ (\overline{\bf
15},{\bf 2})_{-1} + 
({\bf 1},{\bf 1})_{0} +
({\bf 35},{\bf 1})_{0} + 
({\bf 1},{\bf 3})_{0} \nonumber \\ 
&& + ({\bf 15},{\bf 2})_{+1} +
(\overline{\bf 15},{\bf 1})_{+2}+
({\bf 1},{\bf 2})_{+3}\;.
\eea
The embedding tensor, transforming in the ${\bf 912}$ representation,
decomposes into a large number of representations and among them are the 
$(\overline{\bf 6},{\bf 1})_{+4}$ and  
$({\bf 20},{\bf 1})_{+3}$ representations that are potentially 
related to the five- and three-form fluxes of the IIB theory. Hence we
investigate embedding tensors expressed in terms of two tensors,  
$\theta_{\Lambda\Sigma\Gamma}{}^\tau$ and $\theta^\Lambda$, where 
$\Lambda,\Sigma,\ldots= 1,\ldots,6$ and $\tau=1,2$ refer to ${\rm
SL}(6,\mathbb{R})$ and ${\rm SL}(2,\mathbb{R})$ indices in the
defining representations, respectively. These embedding tensors couple
the gauge fields to the 
\Exc7 generators belonging to the $(\overline{\bf 15},{\bf
2})_{-1}+({\bf 15},{\bf 1})_{-2}+({\bf 1},{\bf 2})_{-3}$
representation, which we denote by $t^{\Lambda\Sigma \tau}$,
$t_{\Lambda\Sigma}$ and $t^\tau$, respectively. The generators of the
gauge group are,
\bea
\label{2-theta} 
X_{\Lambda\Sigma\Gamma} &=& 2\,\varepsilon_{\tau\sigma}\,
\theta_{\Lambda\Sigma\Gamma}{}^\tau\, t^\sigma\;, 
\nonumber\\ 
X^{\Lambda\, \tau} &=& \ft16 
\varepsilon^{\Lambda\Sigma\Gamma\Omega\Pi\Delta} \, 
\theta_{\Sigma\Gamma\Omega}{}^\tau \,t_{\Pi\Delta}  + 
 \theta^\Lambda\, t^\tau  \;, 
\nonumber\\[-.6mm]
X_\Lambda &=& \varepsilon_{\tau\sigma}\, 
\theta_{\Lambda\Sigma\Gamma}{}^\tau  \,t^{\Sigma\Gamma \sigma}  +   
\theta^{\Sigma}\,t_{\Lambda\Sigma}\; .
\eea
The above result, which defines the embedding matrix 
$\Theta_M{}^\a$ in terms of $\theta_{\Lambda\Sigma\Gamma}{}^\tau$ and
$\theta^\Lambda$, is uniquely determined by requiring that
$\Theta_M{}^\a$ is an element of the ${\bf 912}$ representation.
Note that there is a certain degeneracy in these definitions; not 
all generators $X_M$ are linearly independent, and there are at most
20 independent generators. 
The quadratic constraint, which ensures the closure of the gauge
generators, implies
\be
\varepsilon^{\Lambda\Sigma\Gamma\Omega\Pi\Delta}
\,\theta_{\Lambda\Sigma\Gamma}{}^\tau\, 
\theta_{\Omega\Pi\Delta}{}^\sigma  =0\;.
\ee
The gauge algebra has the following non-vanishing commutation
relations,   
\bea
\label{gauge-algebra}
{[}X_{\Lambda},X_{\Sigma}]&=& - 2\,
\varepsilon_{\tau\sigma}\,\theta_{\Lambda\Sigma\Gamma}{}^\tau \,
X^{\Gamma \sigma} -
\theta^\Gamma\,
X_{\Lambda\Sigma\Gamma}\;,\nonumber\\[1ex]
{[}X_{\Lambda},X^{\Sigma \tau}]&=& - \ft16
\varepsilon^{\Sigma\Gamma_1\Gamma_2\Gamma_3\Omega_1\Omega_2}\,
\theta_{\Gamma_1\Gamma_2\Gamma_3}{}^\tau\, X_{\Lambda\Omega_1\Omega_2}
\;. 
\eea

Without considering a specific $N=8$ Lagrangian, we have thus constructed
a novel gauging related to a possible IIB flux compactification. The
$T$-tensor and the corresponding potential can be constructed by
choosing a convenient representative for the ${\rm E}_{7(7)}/{\rm
SU}(8)$ coset space. The details can be found in \cite{dWST2}, where
we also prove that gaugings with nilpotent charges lead to a positive
potential. This potential has no stationary points, but for a certain
choice of the embedding tensor one can find domain wall solutions that
can be lifted to ten-dimensional IIB supergravity. 

In the approach of this section, where we simply start from a choice
of the embedding tensor, one avoids the subtleties associated with
electric-magnetic duality. The emphasis is on finding an admissible
embedding tensor, without the need of first constructing the full 
Lagrangian. The approach is equally well applicable to
maximal supergravity in any other number of spacetime dimensions.

\vspace{3mm}
\noindent
This work is partly supported by EU contract HPRN-CT-2000-00131.

\providecommand{\href}[2]{#2}\begingroup\raggedright\endgroup
\end{document}